\documentclass[%
reprint,
 showpacs,
 amsmath,amssymb,
]{revtex4-1}

\usepackage{graphicx}
\usepackage{dcolumn}
\usepackage{bm}

\usepackage{CJK}

\usepackage{color}
\usepackage{amssymb}
\usepackage{amsfonts}

\usepackage[colorlinks,urlcolor=blue,linkcolor=blue,citecolor=blue,anchorcolor=blue]{hyperref}

\begin{document}

\preprint{APS/123-QED}

\title{Hierarchy of Nonlinear Entanglement Dynamics for Continuous Variables}


\author{Da Zhang$^{1,2}$}
\email{zhang1556433@163.com}
\author{David Barral$^2$}
\email{david.barral@lkb.ens.fr}
\author{Yin Cai$^1$}
\email{caiyin@xjtu.edu.cn}
\author{Yanpeng Zhang$^1$}
\author{Min Xiao$^{3}$}
\author{Kamel Bencheikh$^2$}
\email{kamel.bencheikh@c2n.upsaclay.fr}

\affiliation{%
$^1$\mbox{Key Laboratory for Physical Electronics and Devices of the Ministry of Education \& Shaanxi Key Lab of Information }
\mbox{Photonic Technique, School of Electronic and Information Engineering, Xi'an Jiaotong University, Xi'an 710049, China} \\
$^2$\mbox{Centre de Nanosciences et de Nanotechnologies C2N, CNRS, Universit\'e Paris-Saclay, 91120 Palaiseau, France} \\
$^3$\mbox{Department of Physics, University of Arkansas, Fayetteville, Arkansas 72701, USA}
}%

\date{\today}

\begin{abstract}
\noindent
The entanglement produced by a bilinear Hamiltonian in continuous variables has been thoroughly studied and widely used.
In contrast, the physics of entanglement resulting from nonlinear interaction described by partially degenerate high-order Hamiltonians remains unclear.
Here, we derive a hierarchy of sufficient and necessary conditions for the positive-partial-transposition separability of bipartite nonlinear quantum states.
The proposed criteria detect the nonpositive-partial-transposition inseparability of higher-order moments of states, which provides a systematic framework for the characterization of this kind of entanglement.
Through numerical simulation of cubic and quartic Hamiltonians, we demonstrate the existence and competition of a hierarchy of entanglement witnesses, revealing the mechanism underlying such entanglement.
Our results may provide a new direction in continuous variable quantum information processing.
\end{abstract}

\maketitle
A system of two spatially separated quantum particles with maximally correlated momenta and maximally anticorrelated positions is the basis of the thought experiment on the nonlocality of quantum mechanics proposed by Einstein, Podolsky and Rosen in 1935 \cite{einstein.pr.47.777.1935}.
This \emph{innocent} idea gave rise to the concept of quantum entanglement which underpins all current development of quantum technologies \cite{Acn2018}.
Entanglement stands today as a physical resource and its characterization is a primary task as its unveiled properties have great importance for fundamental studies \cite{bell.prf.1.195.1964, reid.rmp.81.1727.2009} and deep impact on quantum information protocols \cite{braunstein2012quantum}.
This characterization is based on well-established criteria which provide various conditions and require different experimental techniques to evaluate the entanglement \cite{guhne.pr.0370.2009,horo.rmp.81.865.2009}.
Determining the existence of entanglement in the laboratory is nevertheless far from being simple, especially for continuous variable systems which involve physical quantities with a continuous spectrum of values, like the position and momentum of particles or the quadratures of the electromagnetic field.

Entanglement is nowadays created regularly in laboratories in different physical systems.
Entangled electromagnetic fields are, for instance, produced in parametric amplifiers and oscillators whose dynamics is described by a bilinear Hamiltonian $\hat{H}_{L}\propto \hat{a}^{\dag}\hat{b}^{\dag}+\mathrm{H.c.}$ \cite{Heidmann.prl.59.2555.1987,zhangda.pra.96.043847.2017}.
The generated states exhibit Gaussian statistics and their entanglement properties are completely characterized by the covariance matrix \cite{vanloock.rmp.77.513.2005}.
Linear correlations as those of $\hat{H}_{L}$ produce \emph{linear entanglement.}
Using the Peres-Horodecki criterion--or positive partial transposition (PPT) criterion--\cite{horodecki.pla.232.1997,peres.prl.77.1413.1998}, Simon derived separability inequalities for Gaussian states whose violation provides a sufficient and necessary condition for entanglement \cite{simon.prl.84.2726.2000}. Other equivalent but different forms of entanglement criteria involving 2nd-order moments were deduced in Refs. \cite{duan.prl.84.2722.2000,mancini.prl.88.120401.2002}.
As an extension of these works, a series of entanglement criteria for multimode Gaussian states were proposed \cite{werner.prl.86.3658.2001,giedke.prl.87.167904.2001,alessio.prl.96.110402.2006,walborn.pra.92.052316.2015,teh.pra.90.062337.2014,teh.pra.100.022126.2019}.
Gaussian states are out of the ordinary because of their simple generation and manipulation, and their versatility in quantum information protocols \cite{vanloock.rmp.77.513.2005,weedrook.rmp.84.621.2012}.

Entanglement can also be created in physical processes described by high-order Hamiltonians such $\hat{H}_{N}\propto \hat{a}^{\dag}\hat{b}^{\dag 2}+\mathrm{H.c.}$ for example \cite{chang.prx.10.011011.2020}, and where one of the operators associated with the entangled modes appear at least at the second power. The classical dynamics of such systems is nonlinear and the quantum statistics of the generated modes is non-Gaussian.
Remarkably, it has been demonstrated that non-Gaussian entangled states can further enhance the performance of quantum information protocols, such as improving the fidelity of teleportation \cite{olivares.pra.67.032314.2003,dell.pra.76.022301.2007} and quantum cloning \cite{cerf.prl.95.070501.2005}, and are also an indispensable component to realize universal quantum computing \cite{l1oyd.prl.82.1784.1999,nielsen.prl.97.110501.2006,eisert.pra.82.042336.2010,eisert.pra.85.062318.2012}.
Because of the nonlinear dynamics nature of the interactions described by high-order Hamiltonians, we refer throughout this Letter to such non-Gaussian entanglement as nonlinear entanglement \cite{lam.prl.114.100403.2015} in order to distinguish it from the non-Gaussian entanglement generated by interactions with bilinear Hamitonians and non-Gaussian operations \cite{masahide.np.4.178.2010}, not the purpose of this Letter.
The usual criteria relying on the computation of the covariance matrix fails to detect such nonlinear entanglement.
Towards the generalized entanglement verification, an infinite hierarchy of conditions based on PPT were proposed in Refs. \cite{vogel.prl.95.230502.2005,Piani.prl.97.058901.2006}.
Although powerful, there is a difficulty in practical application, i.e., how to select an adequate condition to detect entanglement.
Other non-Gaussian entanglement criteria are based on uncertainty relations of constructed special operators \cite{agarwal.njp.211.2005,mm.prl.96.050503.2006,nm.prl.101.130402.2008}, the entropic uncertainty relation  \cite{walborn.prl.103.160505.2009,walborn.pra.83.032307.2011}, and the fidelity of teleportation in quantum channels \cite{Nha.prl.108.030503.2012}.
However some of the proposed non-Gaussian inseparability criteria \cite{walborn.prl.103.160505.2009,walborn.pra.83.032307.2011,Nha.prl.108.030503.2012,ryo.pra.85.062307.2012} are not capable of detecting such nonlinear entanglement, while others are only sufficient conditions \cite{mm.prl.96.050503.2006,nm.prl.101.130402.2008,shchukin.pra.93.032114.2016}.
So far no systematic framework has been provided to study nonlinear entanglement, which leads to its unclear physical connotation.

In this Letter, we propose a hierarchy of sufficient and necessary conditions for the PPT separability of bipartite nonlinear continuous variable quantum states spontaneously generated by the high-order Hamiltonians.
The criteria are applicable for all systems, however in this Letter we focus our analysis on optical systems and the entangled states are optical modes generated in nonlinear materials.
The hierarchy originates from the non-Gaussian statistics of nonlinear entangled states, suggesting that the entanglement is jointly characterized by a set of higher-order covariance matrices.
Our conditions for nonlinear entanglement are based on the PPT of these higher-order covariance matrices.
The proposed criteria allows us to detect the non-PPT inseparability of high-order moments of quantum states produced by a partially degenerate $N$-mode spontaneous parametric downconversion (SPDC), which provides a systematic framework for the study of nonlinear entanglement.
By numerically calculating the final states undergoing the nonlinear interaction in the case of three- and four-mode SPDC, we observe the coexistence and competition among different orders of entanglement witnesses, which further reveals the features of nonlinear entanglement.

Let us start our analysis by considering a bipartite nonlinear system described by the interaction Hamiltonian
\begin{equation}\label{eq1}
\hat{H}^{kl}_I=i\hbar\kappa\hat{a}^{\dag k}\hat{b}^{\dag l}\hat{p}+H.c
\end{equation}
with $k,l \geq 1$ and $k+l\geq3$.
The annihilation operators $\hat{a}$, $\hat{b}$, and $\hat{p}$ describe, respectively, the two down-converted modes $\mathcal{A}, \mathcal{B}$ and the pump mode $\mathcal{P}$.
Let Alice and Bob be, respectively, in possession of modes $\mathcal{A}$ and $\mathcal{B}$.
We define the following nonlinear quadrature operators $\hat{Q}^{nk}_\mathcal{A}=[(\hat{a}^{k})^n+(\hat{a}^{\dag k})^n]/2$ and $\hat{P}^{nk}_\mathcal{A}=i[(\hat{a}^{\dag k})^n-(\hat{a}^{k})^n]/2$ [$\hat{Q}^{nl}_\mathcal{B}=[(\hat{b}^{l})^n+(\hat{b}^{\dag l})^n]/2$ and $\hat{P}^{nl}_\mathcal{B}=i[(\hat{b}^{\dag l})^n-(\hat{b}^{l})^n]/2$] for Alice's (Bob's) subsystem, where $n$ is a positive integer standing for the hierarchy index.
For simplicity, we take now $n=1$ , but the following derivations are general for any $n$.
We gather the nonlinear quadrature operators in the vector $\hat{R}^{kl}=(\hat{Q}^k_\mathcal{A},\hat{P}^k_\mathcal{A},\hat{Q}^l_\mathcal{B},\hat{P}^l_\mathcal{B})^T$ and write the generalized commutation relations as
\begin{align}\label{eq2}
[\hat{R}^{kl}_i,\hat{R}^{kl}_j]=i\Omega^{kl}_{ij},
\end{align}
where $\Omega^{kl} =\text{diag}(J_{k},J_{l})$ and $i,j=1,\dots,4$.
$J_{m}=\text{adiag}(\hat{f}_{m},-\hat{f}_{m})$ is a generalized symplectic matrix that depends on the number of quanta operators $\hat{N}_{\mathcal{A/B}}$ of the involved modes, through the operators $\hat{f}_{m}$ given in the Supplemental Material \cite{nonlinearsm.prl.2020}.

Let's briefly discuss the physical properties in the case of optical states generated by the Hamiltonians (\ref{eq1}) with different initial states.
We consider the partially degenerate three-mode SPDC described by $H^{12}_I$.
This Hamiltonian is nonlinear as the operator $\hat{b}$ is squared.
However, if mode $\hat{a}$ is seeded by a bright coherent state $|\gamma\rangle$, the Hamiltonian under the parametric approximation can be rewritten as $H^{12}_I \simeq i\hbar\gamma^*\kappa_{\alpha_p}\hat{b}^{\dag 2}+\mathrm{H.c}$,
where we have replaced  $\hat{a}$ by the classical complex amplitude $\gamma$, and where $\kappa_{\alpha_p}$ is the product of the pump amplitude $\alpha_p$ and $\kappa$.
It describes single-mode SPDC, where a pump photon is converted into two degenerate photons.
Conversely, if mode $\hat{b}$ is seeded by a bright coherent state $|\beta\rangle$, then $H^{12}_I$ reduces to the well-known bilinear Hamiltonian $H^{12}_I \simeq i\hbar\beta^{*2}\kappa_{\alpha_p}\hat{a}^{\dag}+2i\hbar\beta^{*}\kappa_{\alpha_p}\hat{a}^{\dag}\hat{b}^{\dag}+\mathrm{H.c}$.
In both seeding situations, we end up with bilinear Hamiltonians.
This indicates that seeding alters the statistical properties of the modes--Gaussifies them--and entanglement can be thus detected by Gaussian criteria \cite{kamel.prl.120.043601.2018}.
However, these criteria fails in the spontaneous regime \cite{lam.prl.114.100403.2015}, which is the focus of this work.

We start our analysis, in analogy with the Gaussian states case, by constructing the high-order covariance matrices $V^{kl}$ of quantum states whose dynamics is described by $\hat{H}^{kl}_I$.
The elements of $V^{kl}$ are defined as $V^{kl}_{ij}=\langle\Delta\hat{R}^{kl}_i\Delta\hat{R}^{kl}_j+\Delta\hat{R}^{kl}_j\Delta\hat{R}^{kl}_i\rangle/2$, where $\Delta\hat{R}^{kl}=\hat{R}^{kl}-\langle \hat{R}^{kl}\rangle$ and $\langle \hat{R}^{kl}\rangle=\mathrm{tr}[\hat{R}^{kl}\hat{\rho}]$, with $\hat{\rho}$ being the density operator of the system.
Hence from Eq. (\ref{eq2}) we have $V^{kl}_{ij}+i\langle\Omega^{kl}_{ij}\rangle/2
=\langle\hat{R}^{kl}_i\hat{R}^{kl}_j\rangle-\langle\hat{R}^{kl}_i\rangle\langle\hat{R}^{kl}_j\rangle$.
$V^{kl}$ has an important operational meaning: the entanglement of states produced from bipartite Hamiltonians $\hat{H}^{kl}_I$ may be characterized jointly by a set of $V^{kl}$ because of its non-Gaussian nature.

For states initially in vacuum or thermal states $\langle\hat{R}^{kl}_j\rangle=0$ \cite{nonlinearsm.prl.2020}. These generated states fulfill the following compact statement of the uncertainty principle
\begin{align}\label{eq3}
V^{kl}+\frac{i}{2}\langle\Omega^{kl}\rangle \geq 0.
\end{align}
Every physical state where $\langle\hat{R}^{kl}_j\rangle=0$ must satisfy this relation.
The positivity of the density matrix $\rho$ ensures the validity of inequality (\ref{eq3}).

We now apply the PPT criterion to higher-order covariance matrices \cite{horodecki.pla.232.1997,peres.prl.77.1413.1998}.
The \emph{necessary condition} for separability is that the partially transposed density matrix $\rho^{PT}$ is semi-positive definite.
In continuous-variable language this criterion establishes that a local mirror reflection in the phase space is a symmetry of separable states, such that a separable state after a local mirror reflection should fulfill an uncertainty relation like Eq. (\ref{eq3}) \cite{simon.prl.84.2726.2000}.
A mirror reflection on Bob's subsystem exchanges the operators $\hat{b}$ and $\hat{b}^{\dag}$ such that $\hat{Q}^{l}_\mathcal{B}\rightarrow \hat{Q}^{l}_\mathcal{B}$ and $\hat{P}^{l}_\mathcal{B}\rightarrow -\hat{P}^{l}_\mathcal{B}$.
In terms of the higher-order covariance matrix this results in $\widetilde{V}^{kl}=\Lambda_\mathcal{B} V^{kl}\Lambda_\mathcal{B}$, where $\Lambda_\mathcal{B} =\mathrm{diag}(1,1,1,-1)$. Thus $\widetilde{V}^{kl}$ must satisfy the following uncertainty relation
\begin{align}\label{eq4}
\widetilde{V}^{kl}+\frac{i}{2}\langle\Omega^{kl}\rangle \geq 0.
\end{align}
All separable states must obey this restriction, and its universality is stronger than inequality (\ref{eq3}).
Below, we derive a more useful form based on invariants of higher-order covariance matrices.

By definition $V^{kl}$ is a symmetric matrix, and hence can be written in block form
\begin{align}\label{eq5}
V^{kl}=\left(
    \begin{array}{cc}
      A & C \\
      C^T & B \\
    \end{array}
  \right),
\end{align}
where $A$ and $B$ are the higher-order covariance matrices of the two subsystems corresponding, respectively, to Alice and Bob, while $C$ describes the correlation between them.
Inequality (\ref{eq3}) implies that $A\geq(f_k/2)^2$ and $B\geq(f_l/2)^2$, where $f_m=\mathrm{tr}[\hat{f}_m\hat{\rho}]$.
Williamson's theorem and a suitable singular value decomposition \cite{horn2012matrix,duan.prl.84.2722.2000} allows us to write $V^{kl}$ in standard form \cite{nonlinearsm.prl.2020}
\begin{align}\label{eq6}
V_0=\left(
  \begin{array}{cccc}
    a & 0 & c_1 & 0 \\
    0 & a & 0 & c_2 \\
    c_1 & 0 & b & 0 \\
    0 & c_2 & 0 & b \\
  \end{array}
\right).
\end{align}
The four local symplectic invariants related to $V_0$ are $I_1 =\det A$, $I_2 =\det B$, $I_3 =\det C$ and $I_4 = \det (V_0)$.
The standard form $V_{0}$ enables to recast Eq. (\ref{eq3}) in a unambiguous form of local symplectic invariance
\begin{align}\label{eq7}
&\det(A)\det(B)+[\sqrt{\det(J_k)\det(J_l)}-\det(C)]^2-  \nonumber \\
&\mathrm{tr}[ACBC^T]\geq-\det(B)\det(J_k)-\det(A)\det(J_l).
\end{align}
We apply now a local mirror reflection transformation to the standard form given by Eq. (\ref{eq6}). The invariants $I_1$, $I_2$ and $I_4$ remain unchanged, while $I_{3}=\det(C)$ flips signature.
The uncertainty relation (\ref{eq4}) takes a form similar to Eq. (\ref{eq7}), but with a minus sign in front of $I_3$. Therefore, any separable state with higher-order covariance matrix $V^{kl}$ must obey not only the inequality (\ref{eq3}), but also the more rigorous uncertainty relation (\ref{eq4}), which can be expressed as
\begin{align}\label{eq8}
&\det(A)\det(B)+[\sqrt{\det(J_k)\det(J_l)}-|\det(C)|]^2- \nonumber \\
&\mathrm{tr}[ACBC^T]\geq-\det(B)\det(J_k)-\det(A)\det(J_l).
\end{align}
Inequality (\ref{eq8}), as an invariant under local symplectic transformations, does not depend on the standard form (\ref{eq6}) and is the final form of the necessary condition for the separable states represented by $V^{kl}$.
It is also worth noting that for states with $\det(C)\geq0$, this relation is subsumed by the inequality (\ref{eq7}).
In fact, inequality (\ref{eq8}) is also a sufficient condition for the PPT separability of $V^{kl}$.
In order to show it we formulate the following two lemmas involving the inequalities (\ref{eq7}) and (\ref{eq8}):

\emph{Lemma 1}.-- The state represented by $V^{kl}$ is separable if $\det(V^{kl}-F/2)\geq0$, where $F=\mathrm{diag}(f_k,f_k,f_l,f_l)$.

\emph{Lemma 2}.-- The state described by $V_0$ with $\det C\geq0$ is separable.

These lemmas are demonstrated in the Supplemental Material \cite{nonlinearsm.prl.2020}.
Lemma 1 establishes the sufficient condition for the PPT separability of states represented by higher-order covariance matrices $V^{kl}$, while Lemma 2 establishes the correspondence between this sufficient condition and the quantity $\det(C)$, i.e. quantum states with $\det(C)\geq 0$ fulfill the condition of Lemma 1.
These two lemmas enable us to formulate the following theorem:

\emph{Theorem}.--Inequality (\ref{eq8}) is a sufficient and necessary condition for the PPT separability of $V^{kl}$.

\emph{Proof}.---We consider the two representative examples in turn, $\det C\geq0$ and $\det C<0$.
If $\det C\geq0$, inequality (\ref{eq8}) is absolutely satisfied because it is already covered by Eq. (\ref{eq7}).
According to our Lemmas 1 and 2, $V^{kl}$ is separable.
There are two possible scenarios here for the $\det C<0$.
Once inequality (\ref{eq8}) is violated, $V^{kl}$ is definitely entangled because relation (\ref{eq8}) is a necessary condition for separability.
If inequality (\ref{eq8}) is respected and $\det C<0$, the locally mirror reflected higher-order covariance matrix $\tilde{V}^{kl}$ will present $\det C>0$ due to the signature flip being thus a separable state according to lemma 2.
Since the partial transposed separable state is still separable, this implies the separability of the original state.
This completes the proof of the theorem.

This hierarchy of conditions is based on the PPT criterion which does not detect bound entanglement, i.e., nondistillable entanglement.
Although this effect is rare for continuous variables, as shown in \cite{Horodecki.2003.Bound}, it can indeed be observed not only for Gaussian states \cite{Horodecki.prl.85.2657.2000}, but even for non-Gaussian states \cite{werner.prl.86.3658.2001}.
Note that detecting bound entanglement by a finite number of moments is a hard task even for finite-dimensional systems \cite{horodecki.pra.80.052303.2009}.

Now, regarding nonlinear entanglement, we have the following result:
It is sufficient and necessary to confirm the non-PPT entanglement of a nonlinear bipartite state $\hat{\rho}$ jointly characterized by a series of higher-order covariance matrices $V^{kl}$ if one of them violates inequality (\ref{eq8}).
This statement is the main result of the Letter.

Inequalities (\ref{eq4}) and (\ref{eq8}) are equivalent statements and constitute the PPT criterion at the higher-order covariance matrices level. Thus the violation of $\widetilde{S}^{kl}\geq 0$ with $\widetilde{S}^{kl}\equiv\widetilde{V}^{kl}+i\Omega^{kl}/2$ is also a sufficient and necessary condition of non-PPT entanglement.
It can be rewritten as
\begin{equation}\label{eq9}
\widetilde{\nu}^{kl}_{-}<0,
\end{equation}
with $\widetilde{\nu}^{kl}_{-}\equiv \min(\widetilde{\nu}^{kl})$ and where $\tilde{\nu}^{kl}$ are the symplectic eigenvalues of $\tilde{S}^{kl}$.
Remarkably, a hierarchy of entanglement witnesses $\widetilde{\nu}^{(nk)(nl)}_{-}$ is derived from inequality (\ref{eq9}) for $n\geq1$.
This hierarchy is characterized by the index $n$ which defines the order of the nonlinear correlations.

\begin{figure}[t]
\centering
  \includegraphics[width=7.5cm]{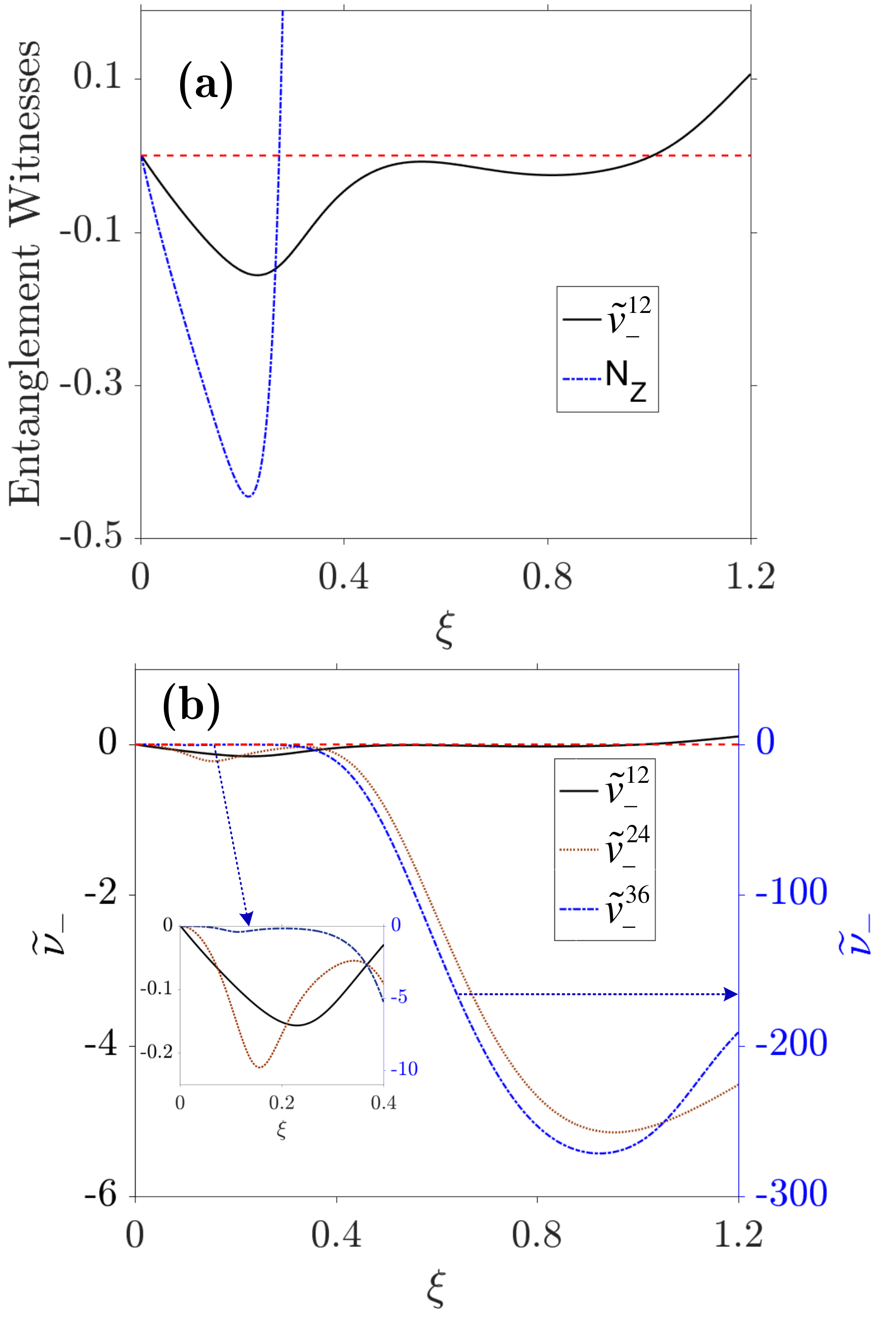}  
  \caption{(a) Comparison of the criteria proposed by Nha and Zubairy and ours. $\widetilde{\nu}^{12}_-$ is the minimum symplectic eigenvalue of $\widetilde{S}^{12}$. (b) Evolution of $\tilde{\nu}_-^{12}$ (3th), $\tilde{\nu}_-^{24}$ (6th) and $\tilde{\nu}_-^{36}$ (9th) versus interaction strength $\xi$ with $\alpha_p=\sqrt{25}$. }
  \label{fig1}
\end{figure}

Next, let us verify the validity of the proposed criteria and compare it with another criterion, in particular the Nha-Zubairy inequality \cite{nm.prl.101.130402.2008}.
We first consider the partially degenerate three-mode SPDC that has been recently observed in a superconducting circuit \cite{chang.prx.10.011011.2020}.
Using the Hamiltonian $\hat{H}^{12}_I$, the master equations are solved numerically to deduce the final state of system at time $t$ considering that the initial state is vacuum for the triplets and a coherent mode $\alpha_p$ for the pump \cite{zhang.pra.013704.2021}.
The Nha-Zubairy inseparability condition can be expressed as  $\mathrm{N_{Z}}\equiv\langle\Delta^2\hat{L}_1\Delta^2\hat{L}_2\rangle-\langle\hat{N}_{\mathcal{B}}+3/4\rangle^2-\langle\Delta\hat{L}_1
\Delta\hat{L}_2\rangle^2<0$ for the triple-photon state, where $\hat{L}_1=\hat{Q}^1_{\mathcal{A}}-\hat{Q}^2_{\mathcal{B}}$ and $\hat{L}_2=\hat{P}^1_{\mathcal{A}}+\hat{P}^2_{\mathcal{B}}$.
The evolution of $\mathrm{N_{Z}}$ versus $\xi=\kappa t\alpha_p$ is shown in Fig. \ref{fig1}(a).
We see that the value of $\mathrm{N_{Z}}$ is lower than 0 when $\xi\leq0.3$, indicating that parties $\mathcal{A}$ and $\mathcal{B}$ are entangled.

\begin{figure}[t]
\centering
  \includegraphics[width=7.5cm]{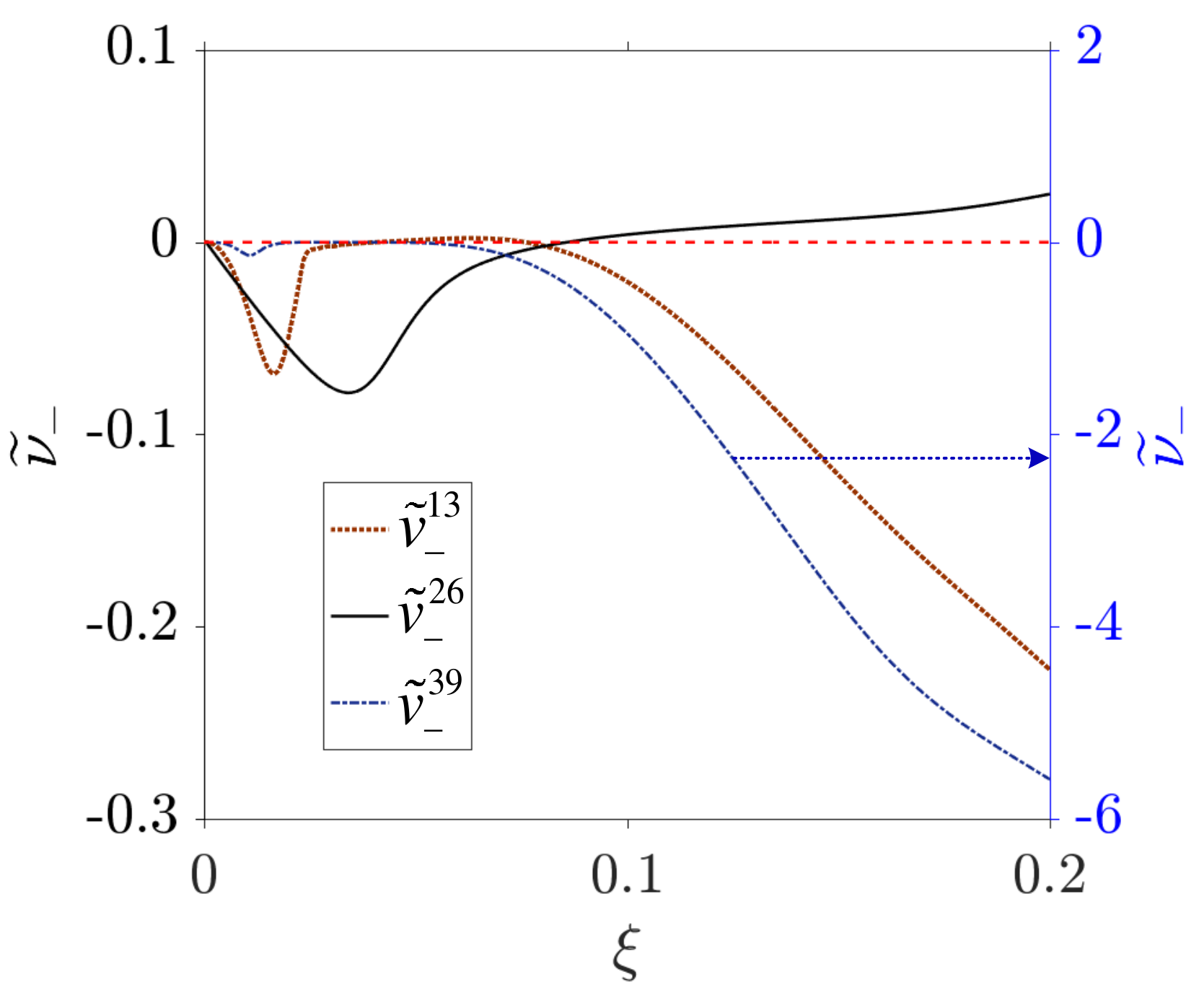}  
  \caption{Evolution of $\tilde{\nu}_-^{13}$ (4th), $\tilde{\nu}_-^{26}$ (8th) and $\tilde{\nu}_-^{39}$ (12th) versus $\xi$ with $\alpha_p=\sqrt{10}$.}
  \label{fig2}
\end{figure}

The lowest-order separability condition related to $\hat{H}^{12}_I$ is $\tilde{S}^{12}\geq 0$ (hierarchy index $n$=1).
In Fig. \ref{fig1}(a), we show the evolution of $\tilde{\nu}_-^{12}$.
The nonlinear entanglement condition (\ref{eq9}) is fulfilled along a wide range of $\xi$.
Note that our criteria has a wider detection range in comparison with the Nha-Zubairy condition.
Interestingly, the entanglement disappears if $\xi\geq1$.
However, from the perspective of the density matrix ${\rho}_{\mathcal{A} \mathcal{B}}$, the entanglement increases with the increase of interaction strength \cite{lam.prl.114.100403.2015}.
This proves that the $V^{12}$ only describes part of the correlated information carried by ${\rho}_{\mathcal{A} \mathcal{B}}$.
Therefore we construct the higher-order covariance matrices $V^{24}$ and $V^{36}$, respectively, with the corresponding separability conditions $\tilde{S}^{24}\geq 0$ ($n$=2) and $\tilde{S}^{36}\geq 0$ ($n$=3).
Figure \ref{fig1}(b) shows the evolution of $\tilde{\nu}_-^{12}$, $\tilde{\nu}_-^{24}$ and $\tilde{\nu}_-^{36}$ versus $\xi$.
The entanglement appears from lower- to higher-order covariance matrices in sequence as the interaction time increases.
More importantly, the hierarchy of entanglement between modes $\mathcal{A}$ and $\mathcal{B}$ coexists over a considerable parameter interval. There could be even higher-order ($n>3$) entanglement that we have not considered.
When $\xi\geq1$, the 3rd-order entanglement between modes $\mathcal{A}$ and $\mathcal{B}$ disappears, and the entanglement transitions to higher-order correlations characterized by 6th- and 9th-order covariance matrices.

As another example of nonlinear entanglement, we consider the four-mode SPDC described by $\hat{H}_I^{13}$.
For such state, we construct the higher-order covariance matrices $V^{13}$, $V^{26}$ and $V^{39}$.
The separability conditions corresponding to inequality (\ref{eq4}) are, respectively, $\tilde{S}^{13}\geq 0$ ($n$=1), $\tilde{S}^{26}\geq 0$, ($n$=2) and $\tilde{S}^{39}\geq 0$ ($n$=3).
The evolution of $\tilde{\nu}_-^{13}$, $\tilde{\nu}_-^{26}$ and $\tilde{\nu}_-^{39}$ are plotted in Fig. \ref{fig2}.
The entanglement of parties $\mathcal{A}$ and $\mathcal{B}$ is mainly carried in the 4th- and 8th-order covariance matrices for small $\xi$.
With the increase of the $\xi$, the entanglement carried by 4th-order covariance matrix disappears, whereas the 8th- and 12th-order dominates.
This is significantly different from the non-Gaussian state prepared by photon addition or subtraction based on the two-mode SPDC, in which the entanglement of the 2nd-order moments--Gaussian--always exists.

There are also other types of competition between the 1st- and 2nd-order coherence of quantum steerable states \cite{kalaga.pra.97.042110.2018}.
We note that the 1st-order coherence does not highlight quantum properties of the field, thus the competition reported in Ref. \cite{kalaga.pra.97.042110.2018} may be widespread in entangled systems.
The separability criteria involving matrices of moments have been derived in several Refs. \cite{horodecki.pra.80.052303.2009,bartkiewicz.pra.91.022323.2015}.
Similarly, we utilize the PPT criterion or other equivalent mappings, but the mechanisms revealed and the detection method provided are completely different.
The matrix of moments composed of creation and annihilation operators are measured by homodyne correlation experiments and some of the criteria proposed in Ref. \cite{horodecki.pra.80.052303.2009} are originally derived from the Cauchy-Schwartz inequality, so even the high-order entanglement criteria cannot reveal the non-Gaussian nature of states.
Reference \cite{bartkiewicz.pra.91.022323.2015} establishes a direct connection between the negativity of the two-qubit state and the higher-order moments.
Other criteria based on probabilities of arbitrary order can be directly applied to photocounting experiments \cite{peifmmode.pra.102.043713.2020}.
These criteria may open new avenues for nonlinear entanglement in discrete variable regime because of their ease of operation.
Our criteria are based on high-order standardized moments measured by heterodyne detection, such as coskewness and cokurtosis \cite{nonlinearsm.prl.2020}.

In summary, we proposed a hierarchy of sufficient and necessary conditions for bipartite nonpositive-partial-transposition entanglement applicable to all nonlinear continuous variable systems described by high-order Hamiltonians (\ref{eq1}).
This provides a systematic framework for the study of this type of nonlinear entanglement.
Subsequently, we observed an interesting phenomenon in nonlinear entangled states, that is, the coexistence and competition between different orders of entanglement witness.
This naturally raises an open question: Can these covariance matrices of different orders be used as independent information encoding or high-dimensional quantum channels in the continuous variable regime?
We expect that the proposed inseparability criteria will help experimentalists to certify nonlinear entanglement features in future experiments.

\section*{Acknowledgement}
The authors thank N. Treps and Z. Zheng for insightful discussions on this study and collaboration.
This work was supported by the National Key Research and Development Program of China (2017YFA0303700, 2018YFA0307500), National Natural Science Foundation of China (61975159, 61605154, 11604256, 11804267, 11904279) and by the Agence Nationale de la Recherche through Project TRIQUI (No. ANR 17-CE24-0041).

%

\end{document}